\documentclass[12pt]{article}

\usepackage{amssymb}
\usepackage{amsthm}
\newtheorem {theorem} {Theorem}%[section]
\newtheorem {definition} [theorem]{Definition}
\newtheorem {proposition} [theorem]{Proposition}

\newtheorem {lemma}  [theorem]{Lemma}
\newtheorem {remark} [theorem]{\bf Remark}

\title{Inverse Jacobi multiplier as a link between \\
conservative systems and Poisson structures}

\author{Isaac A. Garc\'{\i}a$^{\ 1,*}$ and Benito Hern\'andez-Bermejo$^{\ 2}$}

\date{$^{\ (1)}$ {\small Departament de Matem\`atica. Universitat de
Lleida. \\ Avda. Jaume II, 69. 25001 Lleida, Spain.
\\ E--mail: {\tt garcia@matematica.udl.cat}
\\ $ $ \\
$^{\ (2)}$ Departamento de Biolog\'{\i}a y Geolog\'{\i}a, F\'{\i}sica y Qu\'{\i}mica
Inorg\'{a}nica.
\\ Universidad Rey Juan Carlos.
\\ Calle Tulip\'{a}n S/N. 28933--M\'{o}stoles--Madrid, Spain.
\\ E-mail: {\tt benito.hernandez@urjc.es}}}

\begin{document}

\maketitle

\begin{abstract}
Some aspects of the relationship between conservativeness of a dynamical system (namely the preservation of a finite measure) and the existence of a Poisson structure for that system are analyzed. From the local point of view, due to the Flow-Box Theorem we restrict ourselves to neighborhoods of singularities. In this sense, we characterize Poisson structures around the typical zero-Hopf singularity in dimension 3 under the assumption of having a local analytic first integral with non-vanishing first jet by connecting with the classical Poincar\'e center problem. From the global point of view, we connect the property of being strictly conservative (the invariant measure must be positive) with the existence of a Poisson structure depending on the phase space dimension. Finally, weak conservativeness in dimension two is introduced by the extension of inverse Jacobi multipliers as weak solutions of its defining partial differential equation and some of its applications are developed. Examples including Lotka-Volterra systems, quadratic isochronous centers, and non-smooth oscillators are provided.
\end{abstract}


\mbox{}

\mbox{}


% 02.00.00 Mathematical methods in physics
%   02.30.Hq Ordinary differential equations
%   02.30.Ik Integrable systems
% 05.00.00 Statistical physics, thermodynamics, and nonlinear dynamical systems
%   05.45.-a Nonlinear dynamics and chaos
% 45.00.00 Classical mechanics of discrete systems
%   45.20.-d Formalisms in classical mechanics
%   45.20.Jj Lagrangian and Hamiltonian mechanics

\vfill

\footnoterule

$^*$ Corresponding author. Telephone: (+34) 973702728. Fax: (+34) 973702702.

\pagebreak

\section{Introduction}

The presence of finite-dimensional {\it Poisson systems\/} (see \cite{olv1,wei1} and references therein for an overview of the classical theory) is ubiquitous in many branches of physics and applied mathematics. The specific format of Poisson systems has allowed the development of many tools for their analysis (for instance, see \cite{dlrjc1}-\cite{dlrjc3}, \cite{bs6}-\cite{bs8}
and references therein for a sample). In addition, Poisson dynamical systems are significant due to several reasons. One is that they constitute a generalization of classical Hamiltonian systems comprising nonconstant and degenerate structure matrices, as well as odd-dimensional vector fields (in contrast to classical Hamiltonian systems, which are always even dimensional). Additionally, the Poisson system format is not limited by the use of canonical transformations, since every diffeomorphic change of variables maps a Poisson system into another Poisson system.

Let us consider a smooth vector field having a finite-dimensional Poisson structure
\begin{equation}
\label{poisson-V-1}
	\frac{\mbox{\rm d}x}{\mbox{\rm d}t} = {\cal J}(x) \cdot \nabla H (x)
\end{equation}
of dimension $n$ and rank $r \leq n$ constant in a domain (open and simply connected set) $\Omega \subseteq \mathbb{R}^n$. Here ${\cal J}(x)$ and $H(x)$ are the associated structure matrix and Hamiltonian function, respectively. Then under these hypothesis for each point $x_0 \in \Omega$ there is (at least locally in a neighborhood $\Omega_0 \subset \Omega$ of $x_0$) a complete set of functionally independent Casimir invariants $\{ D_{r+1}(x), \ldots , D_n(x) \}$ in $\Omega_0$, as well as a transformation $x \mapsto \Phi(x) = y$ where $\Phi$ is a smooth diffeomorphism in $\Omega_0$ bringing the system (\ref{poisson-V-1}) into its Darboux canonical form. Thus, beyond the fact that Poisson systems are a formal generalization of classical Hamiltonian flows, Darboux Theorem provides the dynamical basis for such a generalization.

{\it Conservative dynamical systems} are those that preserve a finite measure equivalent to a generalized volume. Classical Hamiltonian systems are important examples of conservative systems. Since classical Hamiltonian systems are also a particular case of Poisson systems, it is thus natural that many Poisson systems are also conservative, and conversely that many conservative systems are Poisson systems (but not necessarily Hamiltonian). In spite that the connection between both Poisson systems and conservative flows exists, none of them implies the other, and such link seems to remain relatively unexplored in the literature, at least to the authors' knowledge. The investigation of some aspects of such relationship is the {\it leitmotiv\/} of this work.

More precisely, we say that a $C^1$ vector field $\mathcal{Y} = \sum_{i=1}^n f_i(x) \partial_{x_i}$ defined on $\Omega \subset \mathbb{R}^n$ is conservative if there is a non-negative integrable scalar function $V$ non-identically vanishing on any open subset of $\Omega$ such that the volume integral is preserved under the flow, that is,
\begin{equation}\label{int-inv}
\int_{\Gamma} \frac{d x}{V(x)} = \int_{\varphi_t(\Gamma)} \frac{d x}{V(x)}
\end{equation}
where $\Gamma$ is any measurable subset of $\Omega$ and $\varphi_t(x)$ denotes the associated flow to $\mathcal{Y}$. Various versions of the following result can be found in books such as
\cite{Nemitskii} and \cite{Whittaker}.

\begin{proposition}
The $C^1$ function $V : \Omega \to \mathbb{R}$ which is non-identically vanishing on any open subset of $\Omega$ satisfies (\ref{int-inv}) on any measurable subset $\Gamma \subset \Omega$ if and only if $V$ is a solution of the following linear partial differential equation
\begin{equation}\label{ijm}
\mathcal{Y}(V) = V \, {\rm div} \mathcal{Y},
\end{equation}
where ${\rm div} \mathcal{Y} = \sum_{i=1}^n \partial_{x_i}(f_i(x))$ is the divergence of the $C^1$ vector field $\mathcal{Y}$.
\end{proposition}

Any real $C^1$ non locally null function $V$ in $\Omega$ satisfying (\ref{ijm}) is called {\it inverse Jacobi multiplier}. In 1844 C.G.J. Jacobi introduced in the literature the nowadays called Jacobi (last) multiplier $1/V$. Initially it was mainly used to find the last additional first integral needed to achieve complete integrability of $\mathcal{Y}$. Later, S. Lie found some relationships between $V$ and Lie point symmetries of $\mathcal{Y}$. Recently, it has been proved that the existence of $V$ implies severe consequences to the dynamics of $\mathcal{Y}$ on $\Omega$. In particular, the invariant zero-set $V^{-1}(0)$ contains, under some assumptions, orbits which are relevant in the phase portrait of $\mathcal{Y}$ such as periodic orbits, limit cycles, stable, unstable and center manifolds, etc. (see \cite{BerroneGiacomini2,BGM} for details).
\newline

We will use the following general lemma.

\begin{lemma}\label{lemadivzero}
Any $C^1$ vector field $\mathcal{Y}$ is divergence free if and only if it has the constant inverse Jacobi multiplier $V(x) = 1$.
\end{lemma}
{\it Proof}. Any inverse Jacobi multiplier of $\mathcal{Y}$ satisfies $\mathcal{Y}(V) = V \, {\rm div}(\mathcal{Y})$. Hence it is obvious that if $V(x) = 1$ then ${\rm div}(\mathcal{Y}) \equiv 0$.

Conversely, assume now ${\rm div}(\mathcal{Y}) \equiv 0$. Then any inverse Jacobi multiplier $V$ of $\mathcal{Y}$ satisfies $\mathcal{Y}(V) = 0$, and clearly $V(x) = 1$ is a solution of the equation. $\Box$
\newline

From the point of view of the relationship with Poisson systems and their diffeomorphic transformation properties, it will be useful for us to know how inverse integrating factors change under orbital equivalence of vector fields, see \cite{BerroneGiacomini2,BGM} for further details.

\begin{proposition}\label{V-change}
Let $\Phi$ be a diffeomorphism in $\Omega \subset \mathbb{R}^n$ with non-vanishing Jacobian determinant $J_{\Phi}$ on $\Omega$ and let $\eta : \Omega \to \mathbb{R}$ be such that $\eta \in C^1(\Omega)$ and $\eta(x) \neq 0$ everywhere in $\Omega$. If $V$ is an inverse Jacobi multiplier of the $C^1$-vector field $\mathcal{Y}$ in $\Omega$ then $\eta (V \circ \Phi) / J_{\Phi}$ is an inverse Jacobi multiplier of the orbitally equivalent vector field $\eta \, \Phi_*(\mathcal{Y})$.
\end{proposition}

The structure of the article is the following. Section 2 is devoted to the relationship between three-dimensional conservative and Poisson systems around a zero-Hopf singularity. In Section 3 the concept of strict conservativeness is introduced and its consequences for the existence of a Poisson structure are developed for general $n$-dimensional flows. To conclude, in Section 4 a theory of weak conservativeness for planar flows is outlined.

\section{Characterizing Poisson structures around a zero-Hopf singularity}

In a neighborhood of a regular point, due to the Flow-Box Theorem, any analytic vector field is both Poisson and conservative. Hence, from the local point of view, we shall restrict ourselves to neighborhoods of singular points. In agreement with the result in \cite{agz1}, in this section we shall focus on 3-d Poisson systems that we shall name {\it generic}: given a $3 \times 3$ structure matrix of constant rank 2 in the domain $\Omega \subset \mathbb{R}^3$, such Poisson structure is called generic if there exists one Casimir invariant globally defined in $\Omega$. (Note that in some cases, often related to non-holonomic dynamics, it is possible that a 3-d Poisson structure of constant rank 2 in $\Omega$ is not generic in such domain, see \cite{rus1}-\cite{rus4}, \cite{rus5} and references therein for further details).

The following preliminary result is required:

\begin{lemma}\label{lema-P-R3}
An analytic vector field $\mathcal{Y}$ in an open set $\Omega \subseteq \mathbb{R}^3$ is a generic Poisson system if and only if it is analytically completely integrable in $\Omega$. In such case it can be written as $\mathcal{Y}(x) = \eta(x) \, (\nabla H_2(x) \times \nabla H_1(x))$ where $H_1$ and $H_2$ are independent first integrals and $\eta$ is an inverse Jacobi multiplier of $\mathcal{Y}$. If $\eta$ is a constant then ${\rm div}(\mathcal{Y}) \equiv 0$.
\end{lemma}
{\it Proof}. Clearly in $\Omega \subseteq \mathbb{R}^3$ any generic Poisson system is analytically completely integrable since it possesses two functionally independent analytic first integrals in $\Omega$, namely the Hamiltonian and one Casimir.

Conversely, assume that $\mathcal{Y}$ has two analytic independent first integrals $H_1$ and $H_2$ in $\Omega$. Then it is obvious that it can be written as $\mathcal{Y}(x) = \eta(x) \, (\nabla H_2(x) \times \nabla H_1(x))$ where $x \in \Omega$, $\nabla H_i$ is the gradient of $H_i$, $\eta$ is an analytic scalar function in $\Omega$ and the symbol $\times$ denotes the cross product in $\mathbb{R}^3$. It is straightforward to check that actually such a $\mathcal{Y}$ is a Poisson vector field with Hamiltonian $H_1$ and structure matrix
\begin{equation}\label{structure-J}
{\cal J}(x) = \eta(x) \, \left( \begin{array}{ccc} 0 & \partial_{x_3} H_2(x) & -\partial_{x_2} H_2(x) \\ - \partial_{x_3} H_2(x) & 0 & \partial_{x_1} H_2(x) \\ \partial_{x_2} H_2(x) & - \partial_{x_1} H_2(x) & 0 \end{array} \right).
\end{equation}
Actually $H_2$ becomes the Casimir of ${\cal J}$.

The fact that $\eta(x)$ is an inverse Jacobi multiplier of $\mathcal{Y}$ can be easily checked by direct evaluation.

The last sentence of the lemma follows from ${\rm div}(\nabla H_2(x) \times \nabla H_1(x)) \equiv 0$.  $\Box$

\begin{remark}\label{rem1}
{\rm It is worth emphasizing that the singular points $x_0 \in \Omega \subset \mathbb{R}^3$ where the Poisson vector field vanishes have a special nature. More specifically, since the rank of the structure matrix is assumed to be constant and equal to 2, we have $\mathcal{J}(x_0) \neq 0$. Therefore, we focus on the points where $\nabla H(x_0)=0$, namely on the critical points of the Hamiltonian. In the particular case of $\Omega \subset \mathbb{R}^3$ and after diffeomorphically reducing the system to the Darboux canonical form, it can be seen that the eigenvalues associated to the singularity only can be of the form either $\{ 0 , \pm \lambda \}$ or $\{ 0 , \pm i \omega \}$, with both $\lambda$ and $\omega$ real numbers. In the particular case $\omega \neq 0$ the singularity is called a zero-Hopf singular point. Consequently, this is the generic singularity that can be found in a neighborhood of phase-space completely foliated by periodic orbits.
}
\end{remark}

The previous lemma allows developing the next result.

\begin{theorem}\label{Teo-poisson-V-2}
Let $\mathcal{Y}$ be an analytic vector field in a sufficiently small neighborhood $\Omega \subseteq \mathbb{R}^3$ of a zero-Hopf singularity at $(x_1 ,x_2 ,x_3)=(0,0,0)$ and assume it has an analytic first integral $D$ with $\partial_{x_3} D(0,0,0) \neq 0$ in $\Omega$. Define the 1-parameter family of planar vector fields $\mathcal{Z}_h = \mathcal{Y}|_{\{ D = h \}}$ as the restrictions of $\mathcal{Y}$ to the level sets $\{ D = h \}$ of $D$ with $|h|$ sufficiently small. If $\mathcal{Y}$ is a generic Poisson system in $\Omega$ then $\mathcal{Z}_h$ has a branch of nondegenerate center singularities emerging from the origin at $h=0$. The converse is also true if $\mathcal{Z}_h$ has a family of first integrals depending analytically on $h$.
\end{theorem}
{\it Proof}. Since the origin is a zero-Hopf singularity of $\mathcal{Y}$, its linear part has associated eigenvalues $\{ 0, \pm i \omega \}$ with $i^2 = -1$ and $\omega \in \mathbb{R} \backslash \{0\}$. Performing a linear change of variables and rescaling the time to set $\omega = 1$ we write the linear part of $\mathcal{Y}$ into real Jordan canonical form, that is,
$$
\mathcal{Y} = (-x_2 + F_1(x)) \partial_{x_1} + (x_1 + F_2(x)) \partial_{x_2} + F_3(x) \partial_{x_3}
$$
where the $F_j$ are real analytic functions in $\Omega$ only possessing nonlinear terms. Since the linear part of $\mathcal{Y}$ has two independent first integrals $x_3$ and $x_1^2+x_2^2$ it is clear that $D$ can be chosen in the form $D(x)=x_3 + \cdots$, where the dots denote higher order terms. Then the analytic diffeomorphism $\Phi = ({\rm Id}_2, D)$ in $\Omega$ (where ${\rm Id}_2$ is the identity in $\mathbb{R}^2$) is tangent to the identity and
$$
\Phi_* \mathcal{Y} = (-y_2 + \hat{F}_1(y)) \partial_{y_1} + (y_1 + \hat{F}_2(y)) \partial_{y_2}
$$
where $\hat{F}_i$ are analytic nonlinear terms. By construction it is clear that
$$
\mathcal{Z}_h = (-y_2 + \hat{F}_1(y_1,y_2,h)) \partial_{y_1} + (y_1 + \hat{F}_2(y_1,y_2,h)) \partial_{y_2}
$$
is an analytic family of vector fields defined in a neighborhood of the origin in $\mathbb{R}^2$ and with parameter values of $h$ close to zero. We emphasize that $\mathcal{Z}_h$ has a branch of singularities $(y_1^*(h),y_2^*(h))$ emerging from $(y_1^*(0),y_2^*(0))=(0,0)$ with associated eigenvalues $(\lambda_1(h), \lambda_2(h))$ and $(\lambda_1(0), \lambda_2(0)) = (i, -i)$. Clearly the above singularities are monodromic for $|h|$ sufficiently small.
\newline

We now make use of the assumption that $\mathcal{Y}$ is a generic Poisson vector field in $\Omega$. Then $\mathcal{Y}$ has an analytic first integral $H$ in $\Omega$ functionally independent of $D$. Clearly this additional first integral can be selected as $H(x) = x_1^2+x_2^2 + \cdots$ and the orbits near the origin of $\mathcal{Y}$ are closed since they are the intersection of the level sets of $H$ and $D(x)=x_3 + \cdots$. Thus this $H$ exists if and only if $(y_1^*(h),y_2^*(h))$ is a branch of center singularities of $\mathcal{Z}_h$ and we prove the first part of the theorem.
\newline

Conversely, we assume that $\mathcal{Z}_h$ has the branch $(y_1^*(h),y_2^*(h))$ of nondegenerate center singularities. Therefore $\mathcal{Z}_h$ possesses a family of first integrals $\hat H(x_1, x_2; h) = x_1^2 + x_2^2 + \cdots$ analytic at $(x_1,x_2)=(0,0)$ for any admissible $h$. Furthermore, if additionally $\hat H$ is analytic at $h=0$ then function $H(x) =  \hat H(x_1, x_2; D(x)) = x_1^2 + x_2^2 + \cdots$ is an analytically first integral of $\mathcal{Y}$. Therefore $\mathcal{Y}$ is analytically completely integrable in a sufficiently small neighborhood $\Omega$  and from Lemma \ref{lema-P-R3} it is a Poisson system in $\Omega$. $\Box$

\begin{remark}
{\rm It is interesting to note that from the Poincar\'{e}-Dulac normal form theory there is an analytic diffeomorphism $\Psi$ near the origin such that, when $\mathcal{Z}_h$ has a center at the origin then $\Psi_* \mathcal{Z}_h$ becomes the vector field:
$$
\Psi_* \mathcal{Z}_h = -z_2 (1 + f(z_1^2+z_2^2,h)) \partial_{z_1} + z_1(1 + f(z_1^2+z_2^2,h)) \partial_{z_2}
$$
with $f(0,h)=0$. Moreover, this is a classical Hamiltonian vector field with Hamiltonian function:
$$
    \hat{H}(z_1,z_2;h) = \frac{1}{2} \left( z_1^2 + z_2^2 + \hat{G}( z_1^2 + z_2^2 ; h) \right) \; , \;\: \mbox{with} \:\;\: \hat{G}(w;h) = \int f(w,h) dw.
$$
In short, $\Psi_* \mathcal{Z}_h = (- \partial_{z_2} \hat{H}) \partial_{z_1} + (\partial_{z_1} \hat{H}) \partial_{z_2}$ which is the planar reduction of the Darboux canonical form of the Poisson system $\mathcal{Y}$ of Theorem \ref{Teo-poisson-V-2}.
}
\end{remark}

\begin{remark}
{\rm Observe that, from the Implicit Function Theorem, under the conditions of Theorem \ref{Teo-poisson-V-2} there is an analytic function $\phi(x_1, x_2, h)$ defined in a neighborhood $U$ of the point $(x_1, x_2, h)=(0,0, 0)$ such that $\phi(0, 0, 0)=0$ and satisfies $D(x_1, x_2, \phi(x_1, x_2, h)) \equiv h$ in $U$. Then, since $\mathcal{Y} = (-x_2 + F_1(x)) \partial_{x_1} + (x_1 + F_2(x)) \partial_{x_2} + F_3(x) \partial_{x_3}$, it follows that the reduced vector field $\mathcal{Z}_h$ of Theorem \ref{Teo-poisson-V-2} is $\mathcal{Z}_h = (-x_2 + F_1(x_1, x_2, \phi(x_1, x_2, h))) \partial_{x_1} + (x_1 + F_2(x_1, x_2, \phi(x_1, x_2, h))) \partial_{x_2}$. Clearly in practice we do not have the explicit expression of $\phi$ but we can compute enough terms of the Taylor expansion of $\phi$ at $(x_1, x_2)=(0,0)$. This expansion will permit us to calculate a sufficiently large string of Poincar\'e-Liapunov constants associated to the branch of monodromic nondegenerate singularities $(x_1^*(h), x_2^*(h))$ of the vector field $\mathcal{Z}_h$ in an algorithmic way and try to solve the associated center-focus problem.
}
\end{remark}

\subsection{Example: 3D Lotka-Volterra system}

Let us now consider the quadratic Lotka-Volterra family:
\begin{equation}
\label{lv3de125}
	\left\{ \begin{array}{ccl}
		\dot{x}_1 & = & x_1( \lambda_1 + c x_2 + x_3)    \\
		\dot{x}_2 & = & x_2( \lambda_2 + x_1  + a x_3)   \\
		\dot{x}_3 & = & x_3( \lambda_3 + b x_1 + x_2)
	\end{array} \right.
\end{equation}
These are models of common use in mathematical biology for the description of population interactions. In addition equations (\ref{lv3de125}) are of the Poisson type \cite{Nutku}. Using the Darboux's integrability theory it is easy to check that the full family possesses the inverse Jacobi multiplier $V(x) = x_1 x_2 x_3$. Hence (\ref{lv3de125}) is a conservative family with strictly positive measure in $\Omega = \{ x \in \mathbb{R}^3 : x_i > 0 \}$. On the other hand ${\rm div} \mathcal{Y} \equiv 0$ if and only if $\lambda_1 +\lambda_2 +\lambda_3 =0$ and $a=b=c=-1$. Furthermore (\ref{lv3de125}) has a first integral of the form $I_1(x) = x_1^{1/c} x_2^{b} x_3^{-1}$ when
\begin{equation}\label{param}
	abc=-1 \:\; , \;\:\;\:\;\: \lambda_3 = \lambda_2 b - \lambda_1 ab.
\end{equation}
Now we will use the classical procedure to obtain the additional first integral $I_2(x)$ of (\ref{lv3de125}) using $V(x)$ and $I_1(x)$. More specifically we compute a planar system after restricting (\ref{lv3de125}) to the level sets $\{I_1(x)=h\}$. This can be done substituting $x_3= x_1^{1/c} x_2^{b} / h$ into the two first components of (\ref{lv3de125}) yielding
$$
\dot{x}_1 =  x_1(\lambda_1 + c x_2 + x_1^{1/c} x_2^{b} / h ), \ \ \dot{x}_2 = x_2( \lambda_2 + x_1  + a x_1^{1/c} x_2^{b} / h).
$$
This planar system has the inverse Jacobi multiplier
$$
v(x_1, x_2) = V(x) \partial_{x_3} I_1(x_1, x_2, x_1^{1/c} x_2^{b} / h) = -h x_1 x_2
$$
from where we can obtain the first integral $I(x_1, x_2; h)$ of it having the form
$$
I(x_1, x_2; h) = -x_1 + c x_2 - \frac{a c}{h} x_1^{1/c} x_2^b - \lambda_2 \log x_1 + \lambda_1 \log x_2.
$$
Therefore $I_2(x) = I(x_1, x_2; I_1(x)) = -x_1 + c x_2 - a c x_3 - \lambda_2 \log x_1 + \lambda_1 \log x_2$. We have proved that (\ref{lv3de125}) under the parameter restrictions (\ref{param}) is analytically completely integrable in the domain $\Omega$, hence it is a Poisson system.

In fact we have that the Casimir invariant is $D(x) = I_1(x)$, the Hamiltonian function is
\begin{equation}
\label{lv3de12h}
   H(x) = abx_1+x_2-ax_3+\lambda_3 \ln x_2 - \lambda_2 \ln x_3
\end{equation}
which can be easily deduced from $I_1$ and $I_2$ and the structure matrix is
\[
     {\cal J}(x) = \left( \begin{array}{ccc}
                    0       & cx_1x_2 & bcx_1x_3  \\
                  -cx_1x_2  &   0     & -x_2x_3   \\
                 -bcx_1x_3  & x_2x_3  &  0
              \end{array} \right).
\]

\subsection{Example 1 of Theorem \ref{Teo-poisson-V-2}}

Consider a vector field having a zero-Hopf singularity at the origin. System
\begin{equation}\label{0H-Ej-Lienard-a.1.1}
\begin{array}{lll}
\dot{x}_1 &=& -x_2, \\ \dot{x}_2 &=& x_1 + a x_1^2 + b x_1 x_3, \\ \dot{x}_3 &=& c x_1 x_2 + d x_2 x_3.
\end{array}
\end{equation}
corresponds to case (i) in Theorem 1.4 of \cite{GV}. Note that system (\ref{0H-Ej-Lienard-a.1.1}) has the analytic first integral
\begin{equation}\label{eq:novoH2}
D(x)= x_3 + \cdots = \left\{ \begin{array}{ll}
\frac{c}{d^2} + \left(- \frac{c}{d^2} + \frac{c}{d} x_1 + x_3 \right) {\rm e}^{d x_1}, & \mbox{if $d \neq 0$}, \\
x_3 + \frac{c}2 x_1^2, & \mbox{if $d = 0$}.
\end{array}
\right.
\end{equation}
Therefore $\partial_{x_3} D(0,0,0) = 1 \neq 0$ and the level sets $\{ x \in \Omega : D(x) = h \}$ of $D$ are given by the graph of a function $x_3 = \phi(x_1; h)$. Hence the restriction of the vector field (\ref{0H-Ej-Lienard-a.1.1}) to the level sets $\{ D = h \}$ is $\mathcal{Z}_h$ whose expression is
\begin{equation}\label{0H-Ej-Lienard-a.1.2}
\begin{array}{lll}
\dot{x}_1 &=& -x_2, \\ \dot{x}_2 &=& x_1 + a x_1^2 + b x_1 \phi(x_1; h).
\end{array}
\end{equation}
We note that since $\phi(0; h) = h$, the eigenvalues at the origin of this planar family are $\pm i \sqrt{1+b h}$ and therefore are pure imaginary for small values of $|h|$. Additionally, for such values of $h$, family (\ref{0H-Ej-Lienard-a.1.2}) has a center at the origin because it is a Hamiltonian family. Since the Hamiltonian depends analytically on $h$ near $h=0$ then, using Theorem \ref{Teo-poisson-V-2} we deduce that family (\ref{0H-Ej-Lienard-a.1.1}) has a Poisson structure around the origin.

We will complete this example by explicitly showing the Poisson structure of (\ref{0H-Ej-Lienard-a.1.1}). First we compute the function
$$
\phi(x_1; h)=  \left\{ \begin{array}{ll}
\frac{1}{d^2}(c + {\rm e}^{-d x_1} (-c + d^2 h) - c d x_1), & \mbox{if $d \neq 0$}, \\
h - \frac{c x_1^2}{2}, & \mbox{if $d = 0$}.
\end{array}
\right.
$$
Then the Hamiltonian $\hat{H}(x_1, x_2; h)$ of (\ref{0H-Ej-Lienard-a.1.2}) is
$$
\hat{H}=  \left\{ \begin{array}{ll}
\frac{1}{6 d^4}( {\rm e}^{-d x_1} (-6 b (c - d^2 h) (1 + d x_1) - d^2 {\rm e}^{d x_1} (b c x_1^2 (3 - 2 d x_1) + & \\
 d^2 (x_1^2 (3 + 2 a x_1) + 3 x_2^2))) ) , & \mbox{if $d \neq 0$}, \\
-\frac{1}{24} x_1^2 (12 + 12 b h + 8 a x_1 - 3 b c x_1^2) - \frac{1}{2} x_2^2, & \mbox{if $d = 0$}.
\end{array}
\right.
$$
Now we have the first integral $H(x) = \hat H(x_1, x_2, D(x))$ of (\ref{0H-Ej-Lienard-a.1.1}) given, up to a multiplicative constant, by
$$
H(x)=  \left\{ \begin{array}{ll}
b c (-6 + d^2 x_1^2 (3 + 2 d x_1)) - d^4 (x_1^2 (3 + 2 a x_1) + 3 x_2^2) + & \\
 6 b d^2 (1 + d x_1) x_3, & \mbox{if $d \neq 0$}, \\
-12 x_2^2 - x_1^2 (8 a x_1 + 3 (4 + b c x_1^2 + 4 b x_3)), & \mbox{if $d = 0$}.
\end{array}
\right.
$$
Let $\mathcal{X}$ be the associated vector field to (\ref{0H-Ej-Lienard-a.1.1}). It follows that there exists a scalar function $\eta : \Omega \to \mathbb{R}$ such that $\mathcal{X}(x) = \eta(x) \, (\nabla H(x) \times \nabla D(x))$ where $x \in \Omega$. From the explicit expressions of $D(x)$ and $H(x)$, direct calculations yield
\begin{equation}\label{eq:eta}
\eta(x)=  \left\{ \begin{array}{ll}
-\frac{{\rm e}^{-d x_1}}{6 d^4}, & \mbox{if $d \neq 0$}, \\
- \frac{1}{24}, & \mbox{if $d = 0$}.
\end{array}
\right.
\end{equation}
Finally we obtain that (\ref{0H-Ej-Lienard-a.1.1}) is a Poisson system with Hamiltonian $H$ and structure matrix ${\cal J}(x)$ which comes from (\ref{structure-J}) with $H_2 = D$, that is,
$$
{\cal J}(x) =  - \frac{1}{24} \left( \begin{array}{ccc} 0 & 1& 0 \\-1& 0& c x_1 \\ 0 & -c x_1 & 0 \end{array} \right)
$$
when $d = 0$ and
$$
{\cal J}(x) = \frac{1}{6 d^4} \left( \begin{array}{ccc} 0 & 1& 0 \\ -1& 0 & c x_1 + d x_3 \\ 0 & -c x_1 - d x_3 & 0 \end{array} \right)
$$
if $d \neq 0$.

\subsection{Example 2 of Theorem \ref{Teo-poisson-V-2}}

The quintic family
\begin{eqnarray}
\dot{x}_1 &=& P(x_1, x_2, x_3), \nonumber \\
\dot{x}_2 &=& x_1 +B_2 x_1 x_2 (-x_1^2 + x_3), \label{Ej2} \\
\dot{x}_3 &=& 2 x_1 P(x_1, x_2, x_3) \nonumber
\end{eqnarray}
with $P(x_1, x_2, x_3) = -x_2 -C x_1 x_2 + B_1 (x_1^2 + x_2^2) (-x_1^2 + x_3)$ has a zero-Hopf point at the origin and the first integral $D(x) = x_3- x_1^2$. Then the reduced vector field $\mathcal{Z}_h$ to the level sets $\{ D = h \}$ is given by the planar quadratic family
$$
\begin{array}{lll}
\dot{x}_1 &=& -x_2 - C x_1 x_2 + B_1 h (x_1^2 + x_2^2), \\ \dot{x}_2 &=& x_1 + B_2 h x_1 x_2
\end{array}
$$
having a singularity at the origin with eigenvalues $\pm i$. Therefore, the conditions under which the origin becomes a center for family $\mathcal{Z}_h$ are well known, see the seminal papers \cite{Ka1,Ka2} and \cite{B}. In short, $\mathcal{Z}_h$ has a center for all $h$ if and only if either $B_1=0$ or $C=0$, in which case there is an analytic first integral $\hat H(x_1, x_2; h)$ at $(x_1, x_2)=(0,0)$ for each $h$. We observe that the former center cases are not always Hamiltonian (this situation only appears when $C = 2 B_1 + B_2 =0$). Consequently, if $B_1 C \neq 0$ then family (\ref{Ej2}) does not have a generic Poisson structure in any domain $\Omega \subset \mathbb{R}^3$ containing the origin.
\newline

In the analysis of the first center strata we let $B_1=0$. If $B_2=0$ then $\mathcal{Z}_h = \mathcal{Z}_0$ is independent of $h$ so the first integral $\hat H(x_1, x_2)$ is also independent. When $B_2 \neq 0$ then (see \cite{S})
$$
\hat H(x_1, x_2; h) = (1 + C x_1)^{2 B_2^2 h^2} (1 + B_2 h x_2)^{2 C^2} \, \exp\left[-2 B_2 C h (B_2 h x_1 + C x_2)\right].
$$
The second center strata will be analyzed with parameter restrictions $B_1 \neq 0$ and $C=0$. If moreover $B_2 \neq 0$ then (see again \cite{S})
\begin{eqnarray*}
\hat H(x_1, x_2; h) &=& [-B_2 - B_1 (2 B_1^2 - 3 B_1 B_2 + B_2^2) h^2 x_1^2 - 2 B_1 B_2 h x_2 + \\
 & & B_1^2 (-2 B_1 + B_2) h^2 x_2^2] \, (1 + B_2 h x_2)^{-\frac{2 B_1}{B2}},
\end{eqnarray*}
whereas when $B_2 = 0$ then
\begin{eqnarray*}
\hat H(x_1, x_2; h) &=& (x_1^2 + x_2^2) \, \exp(-2 B_1 h x_2).
\end{eqnarray*}
We note that in any center case the first integral $\hat H(x_1, x_2; h)$ is analytic with respect to $h$ at $h=0$. Therefore, from Theorem \ref{Teo-poisson-V-2} we conclude that family (\ref{Ej2}) has a Poisson structure around the origin if and only if $B_1 C = 0$. In fact, the explicit construction of the Hamiltonian $H(x)$ and structure matrix ${\cal J}(x)$ can be done in a way analogous to that of family (\ref{0H-Ej-Lienard-a.1.1}).

\subsection{Example 3 of Theorem \ref{Teo-poisson-V-2}}

The polynomial family of sixth degree
\begin{eqnarray}
\dot{x}_1 &=& P(x_1, x_2, x_3), \nonumber \\
\dot{x}_2 &=& Q(x_1, x_2, x_3), \label{Ej3} \\
\dot{x}_3 &=& 2 (x_1 P(x_1, x_2, x_3) + x_2 Q(x_1, x_2, x_3)) \nonumber
\end{eqnarray}
with $P(x_1, x_2, x_3) = -x_2 + A x_1^2 + B x_1 x_2 + C x_2^3 + x_1^3 x_3 - x_1^3 x_2^2 - x_1^5$ and $Q(x_1, x_2, x_3) = x_1 + F x_1^2 + E x_2^2 - x_1^3 x_2 - x_1 x_2^3 + x_1 x_2 x_3$  has a zero-Hopf point at the origin and the first integral $D(x) = x_3 - x_1^2-x_2^2$. Then the reduced vector field $\mathcal{Z}_h$ to the level sets $\{ D = h \}$ is given by the planar cubic family
$$
\begin{array}{lll}
\dot{x}_1 &=& R(x_1, x_2; \mu) = -x_2 + A x_1^2 + B x_1 x_2 + h x_1^3 + C x_2^3, \\ \dot{x}_2 &=& S(x_1, x_2; \mu) = x_1 + h x_1^2 + h x_1 x_2 + E x_2^2
\end{array}
$$
having a singularity at the origin with eigenvalues $\pm i$. We have defined the parameter vector $\mu =(h, A, B, C, E) \in I \times \mathbb{R}^4$ where $I \subset \mathbb{R}$ is a small neighborhood of the origin.

We claim that family $\mathcal{Z}_h$ does not have a center at the origin for any parameter value $h \in I$. To prove such claim we will see that the first focal value associated to the origin of $\mathcal{Z}_h$ is not identically zero for any $h \in I$. We briefly recall here the theory of focal values, see for example \cite{RS}. Using the complex coordinate $z = x_1 + i x_2 \in \mathbb{C}$ with $i^2 = -1$, any planar family $\mathcal{Z}_h$ can be written in the form $\dot z = i z + F(z, \bar z; \mu)$ where $\bar{z} = x_1 - i x_2$ and $F(z, \bar z; \mu) = R\left(\frac{1}{2}(z + \bar{z}), \frac{i}{2} (\bar{z} - z); \mu \right) + i S\left(\frac{1}{2}(z + \bar{z}), \frac{i}{2} (\bar{z} - z); \mu \right)$. Finally we complement this complex differential equation with its complex conjugate. Denoting by $w = \bar{z}$ we arrive at the complex polynomial
system
\begin{equation}\label{sist-C^2}
\dot z = i z + F(z, w; \mu), \ \ \dot{w} = - i w + \bar{F}(z, w; \mu).
\end{equation}
Now we define the {\it focus quantities} $g_j(\mu) \in \mathbb{R}[\mu]$ as those polynomials such that $\mathfrak{X}_\mu(\mathcal{H}) = \sum_{j \geq 1} g_j(\mu) (z w)^{j+1}$ where $\mathfrak{X}_\mu = (i z + F(z, w; \mu)) \partial_z + ( - i w + \bar{F}(z, w; \mu)) \partial_{w}$ is the complex vector field in $\mathbb{C}^2$ and $\mathcal{H}(z, w; \mu) = z w + \cdots \in \mathbb{C}[[x,w]]$ is a formal power series. It is known that $\mathcal{Z}_{h^*}$ has a center at the origin for a specific parameter value $\mu = \mu^*$ if and only if $g_j(\mu^*) = 0$ for all $j \in \mathbb{N}$. Performing computations we find that $g_1(\mu) = \frac{1}{4} [(A B) + (3 - 2 A - E) h - h^2]$. Since for every parameter combination we have $g_1(\mu) \not\equiv 0$ for arbitrary $h \in I$ we prove that family $\mathcal{Z}_h$ cannot have a center at the origin for any $h \in I$ and therefore, from Theorem \ref{Teo-poisson-V-2} we know that family (\ref{Ej3}) has no generic Poisson structure around the origin.

\section{Conservativeness and Poisson structure}

First of all, notice that there exist conservative planar vector fields that do not have a Poisson structure and do not describe physically conservative dynamics. For example, the linear system
\begin{equation}\label{consnotPoisson}
\dot{x}_1 = - x_2+ \mu x_1 , \ \dot{x}_2 = x_1 + \mu x_2
\end{equation}
with $\mu \neq 0$, has a focus at the origin and non-negative inverse Jacobi multiplier $V(x_1,x_2)=x_1^2+x_2^2$. Therefore system (\ref{consnotPoisson}) is conservative in $\Omega = \mathbb{R}^2$, which clearly does not correspond to a conservative flow in direct physical terms. Moreover this system is not of Poisson type since it does not have an analytical first integral in any neighborhood of the origin.

This situation is not exceptional. For instance, in \cite{BerroneGiacomini2} the following example appears. System
$$
\dot{x}_1 = \frac{1}{2} [-x_2 + x_1 (1-x_1^2-x_2^2)], \ \dot{x}_2 = \frac{1}{2} [x_1 + x_2 (1-x_1^2-x_2^2)], \ \dot{x}_3 = x_3
$$
has the inverse Jacobi multiplier $V(x) = (x_1^2+x_2^2)^2$ in $\Omega = \mathbb{R}^3$. Since $V$ is non-negative it is clear that the system is conservative in $\mathbb{R}^3$. The eigenvalues of the linearization at the origin are $\{ \frac{1}{2} (1 \pm i), 1 \}$ and therefore, see Remark \ref{rem1}, the system has not a Poisson structure in a neighborhood of the origin. We note that additionally the system also possesses another inverse Jacobi multiplier $V_2(x) = x_3$ which implies the existence of the rational first integral $I_1(x) = V_2(x)/V_1(x) = x_3 / (x_1^2+x_2^2)^2$ not well defined at the origin.
\newline

These examples suggest that the formal definition of conservative flow is not sufficiently restrictive from the most standard physical perspective. However, this difficulty can be overcome in very simple terms just by imposing that the Jacobi multiplier be strictly positive (or strictly negative, since $-V$ is an inverse Jacobi multiplier if and only if $V$ is). It is clear that system (\ref{consnotPoisson}) has a positive inverse Jacobi multiplier if and only if $\mu = 0$ and thus the origin is a center and the flow becomes Hamiltonian. This simple example reflects what is going to be the actual general situation in the two-dimensional case.

\begin{definition}
{\rm A conservative vector field is {\em strictly conservative} if it has a strictly positive invariant measure, that is, satisfying (\ref{int-inv}) with $V > 0$ in $\Omega$.}
\end{definition}

The trivial examples of strictly conservative systems are the divergence free systems, see Lemma \ref{lemadivzero}. The next theorem connects the property of being strictly conservative with the existence of a Poisson structure depending on the phase space dimension.

\begin{theorem}\label{Teo-Poisson-Conservative}
Let $\mathcal{Y}$ be a smooth vector field in $\Omega \subseteq \mathbb{R}^n$. Then:
\begin{itemize}
    \item[(i)]  In the case $n=2$, if $\mathcal{Y}$ is strictly conservative with smooth invariant measure in a simply-connected domain $\Omega$ then it is a Poisson vector field in $\Omega$.

    \item[(ii)] Let $V$ be an inverse Jacobi multiplier of $\mathcal{Y}$ in $\Omega \subseteq \mathbb{R}^2$. Then the zero-set $V^{-1}(0) \subset \Omega$ is an invariant curve that induces a natural partition of $\Omega$ into $m$ disjoint invariant domains $\Omega_i$ with boundaries $\partial \Omega_i \subset V^{-1}(0)$ such that $\cup_{i=1}^m \partial \Omega_i = V^{-1}(0)$. Provided $\Omega_i$ is simply connected, the restricted field  $\mathcal{Y}|_{\Omega_i}$ is a Poisson system orbitally equivalent to the Darboux canonical form which can be constructed globally in $\Omega_i$.

    \item[(iii)] In the case $n = 3$, if $\mathcal{Y}$ is strictly conservative in $\Omega$ it is not necessarily a generic Poisson vector field in $\Omega$.
\end{itemize}
\end{theorem}
{\it Proof}. In the planar case (i), if the smooth vector field $\mathcal{Y} = P(x_1,x_2) \partial_{x_1} + Q(x_1,x_2) \partial_{x_2}$ is strictly conservative in $\Omega \subset \mathbb{R}^2$ with a smooth invariant measure given by $d x_1 d x_2/ V(x_1, x_2)$ then we can construct a first integral $H$ of $\mathcal{Y}$ in $\Omega$ as the line integral
\begin{equation}\label{H}
H(x_1,\,x_2) = \int_{(x_1^0,\,x_2^0)}^{(x_1,\,x_2)} \frac{P(x_1,\,x_2)\,dx_2-Q(x_1,\,x_2)\,dx_1}{V(x_1,\,x_2)}
\end{equation}
along any curve connecting an arbitrarily chosen point $(x_1^0,\,x_2^0)$ and the point $(x_1,\,x_2)$ in $\Omega$. We remark that this line integral might not be well-defined if $\Omega$ is not simply-connected, which is not our case. Also, clearly, since $V > 0$ and smooth in $\Omega$ we find that $H$ is smooth in $\Omega$, and consequently $\mathcal{Y}$ is a Poisson vector field in $\Omega$ (see \cite{CF} for the last sentence). This proves statement (i).
\newline

The proof of (ii) is constructive. First we recall that since $V$ satisfies $\mathcal{Y}(V) = V \, {\rm div} \mathcal{X}$ it is obvious that the zero-set $V^{-1}(0) \subset \Omega$ is an invariant curve of $\mathcal{Y}$. Therefore the induced partition $\{ \Omega_i \}_{i=1}^m$ of $\Omega$ is formed by disjoint invariant domains $\Omega_i$ with boundaries $\partial \Omega_i \subset V^{-1}(0)$ and $\cup_{i=1}^m \partial \Omega_i = V^{-1}(0)$.

In what follows we restrict the analysis to one single simply connected $\Omega_i$. Since $\Omega_i \not\subset V^{-1}(0)$ the planar vector field $\mathcal{Y} = P(x_1,x_2) \partial_{x_1} + Q(x_1,x_2) \partial_{x_2}$ is strictly conservative in $\Omega_i$. Then it follows that ${\rm div}(\mathcal{Y} / V) \equiv 0$ in the set $\Omega_i$. Since $\Omega_i$ is simply connected, there exists a smooth function $H : \Omega_i \to \mathbb{R}$ given by (\ref{H}) such that $\mathcal{Y} / V$ is a Hamiltonian vector field in $\Omega_i$ with Hamiltonian $H$. This induces the noncanonical Poisson structure of $\mathcal{Y}$ in $\Omega_i$ in terms of Hamiltonian $H$ and structure matrix
\begin{equation}\label{J-V}
{\cal J}(x_1,x_2) = V(x_1, x_2) \left( \begin{array}{cc} 0 & 1 \\ -1 &  0 \end{array} \right).
\end{equation}
This completes the proof of (ii).
\newline

To prove (iii) a counterexample can be used in dimension $n = 3$. We will show a vector field $\mathcal{Y}$ in a sufficiently small neighborhood $\Omega \subset \mathbb{R}^3$ of a zero-Hopf singularity at the origin such that $\mathcal{Y}$ is strictly conservative in $\Omega$ but it is not a generic Poisson vector field in $\Omega$. Let us consider the quadratic family of vector fields in $\mathbb{R}^3$
\begin{equation}\label{0H-Ej-1}
\begin{array}{lll}
\dot{x}_1 &=& -x_2, \\ \dot{x}_2 &=& f(x_1,x_3) + x_2 g(x_1,x_3), \\ \dot{x}_3 &=& F(x_1,x_2,x_3),
\end{array}
\end{equation}
where $f(x_1, x_3) = x_1 + a_0 x_1^2 + a_1 x_1 x_3 + a_2 x_3^2$, $g(x_1, x_3) = b_0 x_1 + b_1 x_3$ and $F(x_1, x_2, x_3) = c_0 x_1^2 + c_1 x_2^2 + c_2 x_3^2 + c_3 x_1 x_2 + c_4 x_1 x_3 + c_5 x_2 x_3$ being $a_i, b_i, c_i \in \mathbb{R}$ the parameters of the family. It can be seen that ${\rm div}(\mathcal{Y}) \equiv 0$ and consequently $\mathcal{Y}$ is strictly conservative in $\Omega$ (see Lemma \ref{lemadivzero}) if and only if
\begin{equation}\label{conddiv0}
b_0+c_4 = b_1 + 2 c_2 = c_5 = 0.
\end{equation}

We consider $\hat{\mathcal{Y}} = -x_2 \partial_{x_1} +(f(x_1,x_3) + x_2 g(x_1,x_3)) \partial_{x_2} + F(x_1,x_2,x_3) \partial_{x_3}$, the three-dimensional vector field formed by the first components of (\ref{0H-Ej-1}). Using Theorem 1.5 of \cite{GV}, we know that there is a neighborhood $\hat\mathcal{U}$ of the origin in $\mathbb{R}^3$ completely foliated by periodic orbits of $\hat{\mathcal{Y}}$, including continua of equilibria as trivial periodic orbits, if and only if one of the following parameter conditions hold:
\begin{itemize}
\item[(A)] $f(x_1,x_3) = x_1 + a_0 x_1^2 + a_1 x_1 x_3$, $g(x_1, x_3) \equiv 0$ and $F(x_1,x_2,x_3) = c_3 x_1 x_2 + c_5 x_2 x_3$;

\item[(B)] $f(x_1,x_3) =x_1 + a_0 x_1^2 + a_2 x_3^2$, $g(x, z) \equiv 0$ and $F(x_1,x_2,x_3) = c_3 x_1 x_2 + c_5 x_2 x_3$;

\item[(C)] $f(x_1,x_3) =x_1$, $g(x_1, x_3) = b_0 x_1$ and $F(x_1,x_2,x_3) = c_0 x_1^2 + c_3 x_1 x_2 - c_0 x_2^2 + c_4 x_1 x_3 + c_5 x_2 x_3$ with the parameter restriction $b_0 c_0 c_4 - c_0 c_4^2 - c_3 c_4 c_5 + c_0 c_5^2 = 0$;

\item[(D)] $f(x_1,x_3) =x_1 + a_1 x_1 x_3$, $g(x_1, x_3) = b_0 x_1$ and $F(x_1,x_2,x_3) = c_3 x_1 x_2 + c_4 x_1 x_3$;

\item[(E)] $f(x_1,x_3) = x_1 + a_1 x_1 x_3 + a_2 x_3^2$, $g(x_1, x_3) \equiv 0$ and $F(x_1,x_2,x_3) = c_3 x_1 x_2 + c_5 x_2 x_3$.
\end{itemize}
On the other hand, in \cite{Isaac} it is proved that $\hat\mathcal{U}$ exists if and only if $\hat{\mathcal{Y}}$ is completely analytically integrable, that is, there are two independent analytic first integrals in $\hat\mathcal{U}$. Now it is easy to check that there are vector fields $\hat{\mathcal{Y}}$ satisfying (\ref{conddiv0}) that do not satisfy any of the conditions (A-E). The counterexample follows by taking one of these vector fields and $\hat\Omega \subset \hat\mathcal{U}$, since generic Poisson vector fields in $\hat\mathcal{U}$ are in particular completely analytically integrable in $\hat\mathcal{U}$. This proves part (iii). This completes the proof of statement (iii). $\Box$

\begin{remark}\label{Rem-*}
{\rm Let $x^* \in \partial \Omega_i \subset V^{-1}(0)$ be a point of the boundary of $\Omega_i$. Observe that the rank of the Poisson structure matrix (\ref{J-V}) vanishes on $x^*$. Accordingly, the Darboux canonical form is not defined on such boundary, since there is no constant-rank neighborhood of $x^*$. At the same time, it is worth recalling that the line integral defining the Hamiltonian $H$ in (\ref{H}) is not defined on any neighborhood of $x^*$ due to the vanishing of $V$ on such point.}
\end{remark}

\begin{remark}
{\rm Under the same hypotheses of statement (ii) of Theorem \ref{Teo-Poisson-Conservative}, if some $\Omega_i$ is not simply connected then the same construction is valid for every simply connected subdomain of it. This implies that $\Omega_i$ can be fully decomposed as the union of simply connected subdomains on which the Darboux canonical form can be globally constructed. }
\end{remark}

\subsection{Example: Poisson structure of the quadratic isochronous centers}

An isolated singular point of $\mathcal{Y}$ is said to be a {\it center} if every orbit in a punctured neighborhood of it is a periodic orbit. Additionally, it is said to be an {\it isochronous center} if every periodic orbit in such a neighborhood has the same period. For the class of planar quadratic vector fields having an isochronous center at the origin, Loud proved in \cite{L} that after a linear change of coordinates and a constant time rescaling the system can be brought into four canonical forms. Use of statements (i) and (ii) of Theorem \ref{Teo-Poisson-Conservative} can be made in a similar way on the four isochronous cases in order to construct their Poisson structure and invariant measures. For the sake of illustration only one of them will be analyzed here. For this purpose the following isochronous system $\mathcal{Y}$ is chosen:
\begin{equation}\label{Isoc-1}
\dot{x}_1 = - x_2 - \frac{4}{3} x_1^2, \ \dot{x}_2 = x_1 \left(1- \frac{16}{3} x_2 \right).
\end{equation}
In \cite{CS} it is found the inverse Jacobi multiplier $V(x_1, x_2) = (3 - 16 x_2) (9 - 24 x_2 + 32 x_1^2)$. The set $V^{-1}(0)$ is composed by one straight line and one parabola which do not intersect. Thus $\Omega = \mathbb{R}^2$ has the natural partition given by $\Omega_1 = \{ (x_1, x_2) \in \Omega : 9 - 24 x_2 + 32 x_1^2 < 0 \}$, $\Omega_2 = \{ (x_1, x_2) \in \Omega : 3 - 16 x_2 < 0 < 9 - 24 x_2 + 32 x_1^2  \}$ and $\Omega_3 = \{ (x_1, x_2) \in \Omega : 3 - 16 x_2 > 0 \}$. From the previous discussion it is found that (\ref{Isoc-1}) is a Poisson vector field in each $\Omega_i$ with the same structure matrix
$$
{\cal J}(x_1,x_2) = (3 - 16 x_2) (9 - 24 x_2 + 32 x_1^2) \left( \begin{array}{cc} 0 & 1 \\ -1 &  0 \end{array} \right).
$$
and Hamiltonian
$$
H(x_1, x_2) = \frac{1}{384} \log \frac{|-3 + 16 x_2|}{(18 + 64 x_1^2 - 48 x_2)^2}
$$
obtained after evaluation of the line integral (\ref{H}). As anticipated in Remark \ref{Rem-*}, the structure matrix ${\cal J}$ becomes singular on $V^{-1}(0) = \cup_{i=1}^3 \partial \Omega_i$ and, in addition, $H$ is smooth on each $\Omega_i$ but it is not defined on $V^{-1}(0)$.

\begin{remark}
\label{remark13}
{\rm Let $\mathcal{Y}$ be a smooth Poisson vector field in $\Omega \subseteq \mathbb{R}^n$ with constant rank $r$. Then for every $x_0 \in \Omega$ there is a neighborhood $\Omega_0 \subset \Omega$ of $x_0$ and a smooth diffeomorphism $\Phi$ in $\Omega_0$ such that $\Phi_* \mathcal{Y}$ is written in the Darboux canonical form, hence as a Poisson system with symplectic structure matrix $\mathcal{S}_{(n,r)}$ of dimension $n$ and rank $r$. Since Darboux theorem is not constructive, sometimes in practice \cite{bs6}-\cite{bs8} only a more general structure matrix $\eta \, \mathcal{S}_{(n,r)}$ can be reached for $\Psi_* \mathcal{Y}$ under a different diffeomorphism $\Psi$. In this case an additional time rescaling is required to complete the Darboux reduction. Now it is easy to check that function $\eta : \Omega_0 \to \mathbb{R}$ is just an inverse Jacobi multiplier of the Darboux canonical form $(1/\eta) \Psi_* \mathcal{Y}$. This implies that $V = \eta /J_{\Psi}$ is an inverse Jacobi multiplier of the original Poisson system $\mathcal{Y}$ according to Proposition \ref{V-change}. }
\end{remark}

\section{A theory of weak conservativeness}

Non-smooth differential systems are natural models in many branches of science such as mechanics, electromagnetic theory, automatic control, etc (see for instance \cite{BBCK, F, Kunze}). Therefore we conclude by generalizing the previous planar theory to vector fields with less regularity. The first step is to extend the definition of inverse Jacobi multiplier for a planar vector field $\mathcal{Y}$ as weak solutions of the partial differential equation (\ref{ijm}). In this section we will restrict ourselves to $C^1$ vector fields ${\cal Y} = P(x,y) \partial_x + Q(x,y) \partial_y$ defined on simply connected domains $\Omega \subset \mathbb{R}^2$ having smooth boundary $\partial \Omega$.

The well known test functions will be used to introduce the forthcoming Definition \ref{dwifi3} which is our definition of weak solution of the partial differential equation (\ref{ijm}) defining the classical (hence $C^1$) inverse Jacobi multipliers. Recall that a function $\varphi: \Omega \to \mathbb{R}$ is called {\it test function} if $\varphi \in C^1(\Omega)$ and there is a compact set $K \subset \Omega$ such that the support of $\varphi$ is included in $K$. The linear space of all the test functions in $\Omega$ is denoted by ${\cal D}(\Omega)$.

\begin{definition}\label{dwifi3}
{\rm A function $W: \Omega\subset\mathbb{R}^2 \to \mathbb{R}$ is a {\it weak inverse Jacobi multiplier} in $\Omega$ of the vector field ${\cal Y} = P(x,y) \partial_x + Q(x,y) \partial_y$ with integrable divergence in $\Omega$ provided $W$ is integrable in $\Omega$ and verifies
$$
\int_\Omega W \, [ \mathcal{Y}(\varphi) + 2 \varphi \; {\rm div} \mathcal{Y} ] \ dx dy = 0,
$$
for all $\varphi \in {\cal D}(\Omega)$.}
\end{definition}

The next result gives a relationship between inverse Jacobi multipliers in the plane and weak inverse Jacobi multipliers.

\begin{theorem}\label{twifi1}
Let $V: \Omega\subset\mathbb{R}^2 \to \mathbb{R}$ and $\mathcal{Y}$ a vector field in $\Omega$ both $C^1(\Omega)$. Then $V$ is a weak inverse Jacobi multiplier of $\mathcal{Y}$ in $\Omega$ if and only if it is an inverse Jacobi multiplier.
\end{theorem}
{\it Proof.}
Let $V$ be an inverse Jacobi multiplier of ${\cal Y}$ in $\Omega$. Consequently we have
\begin{equation}
\int_\Omega [\mathcal{Y}(V) - V \, {\rm div} \mathcal{Y}] \ \varphi \ dx dy = 0 \ ,
\label{wifi2}
\end{equation}
for every $\varphi \in {\cal D}(\Omega)$. Now taking into account the identities
$$
P V_x \varphi = (P V \varphi)_x - (P \varphi)_x V \ , \ \ Q V_y \varphi = (Q V \varphi)_y - (Q \varphi)_y V \ ,
$$
the integrand of (\ref{wifi2}) can be rewritten in the form
\begin{eqnarray*}
[\mathcal{Y}(V) - V {\rm div} \mathcal{Y} ] \ \varphi & = & (P V \varphi)_x + (Q V \varphi)_y -  V \{ \varphi \; {\rm div} \mathcal{Y} + {\rm div} [\varphi \mathcal{Y} ] \} \\
 & = & {\rm div} [ V \varphi \mathcal{Y} ] - V \{ \varphi \; {\rm div} \mathcal{Y} - {\rm div} [\varphi \mathcal{Y}] \} \ .
\end{eqnarray*}
Due to the additivity of the integral, equation (\ref{wifi2}) becomes
\begin{equation}
\int_\Omega {\rm div} [ V \varphi \mathcal{Y} ] \ dx dy - \int_\Omega V \{ \varphi \; {\rm div} \mathcal{Y} + {\rm div} [\varphi \mathcal{Y}] \ dx dy = 0 \ ,
\label{iiisaac}
\end{equation}
for every $\varphi \in {\cal D}(\Omega)$. The first integral of the previous expression vanishes. To see this, Green's theorem on the plane can be applied as follows
$$
\int_\Omega {\rm div} [ V \varphi \mathcal{Y} ] \ dx dy = \int_{\partial \Omega} - V \varphi Q \ dx +  V \varphi P \ dy = \int_{\partial \Omega} V \varphi [ P \ dy - Q \ dx ] = 0 \ ,
$$
where in the last step we have used the fact that $\varphi \in {\cal D}(\Omega)$ and therefore $\varphi(x,y)=0$ for all $(x,y) \in \partial\Omega$.
To conclude, after rearrangement of the second integral in (\ref{iiisaac}), equation (\ref{iiisaac}) becomes
$$
\int_\Omega V [ \mathcal{Y}(\varphi) +2 \varphi \; {\rm div} \mathcal{Y} ] \ dx dy = 0 \ .
$$
Then $V$ is a weak inverse Jacobi multiplier after Definition \ref{dwifi3}, thus concluding the proof in one sense.

Now if $V$ is $C^1(\Omega)$ then the converse holds by reversing the previous steps. The proof is thus complete. $\Box$

\subsection{Piecewise $C^k$ weak inverse Jacobi multipliers}

The following definition is now introduced.
\begin{definition}\label{defwiifisaac}
{\rm Let $\Omega\subset\mathbb{R}^2$ be a domain with smooth boundary $\partial\Omega$. A function $f: \Omega \to \mathbb{R}$ is termed {\rm piecewise $C^k$} with $k \geq 1$ in $\Omega$ if there exists a smooth curve $\gamma \subset \Omega$ such that $\Omega_1 \cup \Omega_2 \cup \gamma = \Omega$, where $\gamma =\partial \Omega_1 \cap \partial \Omega_2$ and $f$ verifies $f \in C^k(\Omega_1 \cup \Omega_2)$, but $f \not\in C^1(\gamma)$. Additionally, a planar vector field $\mathcal{Y}$ is piecewise $C^k$ in $\Omega$ if both components are piecewise $C^k$ in $\Omega$ with respect to the same curve $\gamma$.   }

\end{definition}

The forthcoming theorem will be useful in the context of weak inverse Jacobi multipliers.

\begin{theorem}\label{Teo-inv}
Let $\mathcal{Y}$ be a planar vector field admitting a weak inverse Jacobi multiplier $W$ piecewise $C^1$ in $\Omega$ for the curve $\gamma \subset \Omega$ and $W \not\in C(\gamma)$. Then $\gamma$ is an invariant curve of $\mathcal{Y}$.
\end{theorem}
{\it Proof.} Since $W$ is a weak inverse Jacobi multiplier for ${\cal Y}$ in $\Omega$ we have
\begin{equation}\label{int_suma}
\int_\Omega W [ \dot{\varphi} + 2 \varphi  \; {\rm div} \mathcal{Y} ] \ dx dy =
\sum_{i=1}^2 \int_{\Omega_i} W_i [ \dot{\varphi} + 2 \varphi  \; {\rm div} \mathcal{Y} ] \ dx dy = 0 \ ,
\end{equation}
for all $\varphi \in {\cal D}(\Omega)$, where $W_i = W|_{\Omega_i}$. Since $W_i \in C^1(\Omega_i)$, we can make use of the identity
$$
W_i \varphi \; {\rm div} \mathcal{Y} = {\rm div} [ W_i \varphi \mathcal{Y} ] - \varphi \mathcal{Y}(W_i) - W_i \mathcal{Y}(\varphi) \ .
$$
Since $\mathcal{Y}(W_i) = W_i {\rm div} \mathcal{Y}$ in $\Omega_i$, the previous identity can be written in $\Omega_i$ as
$$
2 W_i \varphi \; {\rm div} \mathcal{Y} = {\rm div} [ W_i \varphi \mathcal{Y} ] - W_i \mathcal{Y}(\varphi) \ .
$$
This allows writing (\ref{int_suma}) as
\begin{equation}
\sum_{i=1}^2 \int_{\Omega_i} {\rm div} [ W_i \varphi \mathcal{Y} ] \ dx dy = 0 \ .
\label{wiifinter1}
\end{equation}
Applying Green's theorem to the two previous integrals we have
$$
\int_{\Omega_i} {\rm div} [ W_i \varphi \mathcal{Y} ] \ dx dy = \int_{\partial\Omega_i} W_i \varphi (P \ dy - Q \ dx) = (-1)^{i+1} \int_{\gamma} W_i \varphi (P \ dy - Q \ dx) \ ,
$$
where in the last step we use that by definition of test function in $\Omega$ it is $\varphi(x,y)=0$ for all $(x,y) \in \partial\Omega_i \backslash \gamma$ for $i=1,2$. The factor $(-1)^{i+1}$ takes into account the fact that the line integral has an opposite sense of integration for $\partial\Omega_1$ and $\partial\Omega_2$. Therefore condition (\ref{wiifinter1}) yields
$$
\int_{\gamma} (W_1-W_2) \varphi (P \ dy - Q \ dx)  = 0,
$$
for all $\varphi \in {\cal D}(\Omega)$. Since $W \not\in C(\gamma)$ we conclude that $P \ dy - Q \ dx = 0$ on $\gamma$. Consequently $\gamma$ is an invariant curve of $\mathcal{Y}$. $\Box$

\begin{remark}\label{rem}
{\rm Some simple examples of weak inverse Jacobi multiplier $W$ of ${\cal Y}$ in $\Omega$ are listed below:
\begin{itemize}
\item[(i)] Bounded piecewise $C^1$ in $\Omega$ functions $W$ with respect to a curve $\gamma \subset \Omega$. If additionally $W|_{\Omega_i} > 0$ then $1/W$ is an invariant measure for ${\cal Y}$ in $\Omega$.

\item[(ii)] Consider an inverse Jacobi multiplier $V$ of ${\cal Y}$ in $\Omega$ and assume there is a curve $\gamma = V^{-1}(0)$ that induces a partition $\Omega =\Omega_1 \cup \Omega_2 \cup \gamma$, where $\gamma =\partial \Omega_1 \cap \partial \Omega_2$, with $V|_{\Omega_i}$ having opposite signs. Then $W = |V|$ is a weak inverse Jacobi multiplier of ${\cal Y}$ in $\Omega$.
\end{itemize}
 }
\end{remark}

\subsection{Example: Perturbing a non-smooth harmonic oscillator}

As an instance of statement (i) of Remark \ref{rem}, consider the mechanical model of the non-smooth harmonic oscillator $\ddot y + 2 \, {\rm sign}(y) = 0$, see \cite{BCT}. Taking $\dot{y} = 2 x$, the associated piecewise smooth vector field $\mathcal{Y}_0$ in the $(x,y)$-phase plane is given by $\mathcal{Y}_0^+$ if $y >0$ and $\mathcal{Y}_0^-$ when $y <0$ where $\mathcal{Y}_0^\pm = (\mp 1) \partial_x + 2 x \partial_y$. Now we shall perturb $\mathcal{Y}_0$ as follows: $\mathcal{Y}^\pm_\varepsilon = (\mp 1) \partial_x + (2 x + \varepsilon y) \partial_y$. Define the semi-planes $\Omega^{+} = \{ (x,y) \in \mathbb{R}^2 : y \geq 0 \}$ and $\Omega^{-} = \{ (x,y) \in \mathbb{R}^2 : y < 0 \}$. It is direct to check that $V^\pm(x,y) = \exp(\mp \varepsilon x)$ is an inverse Jacobi multiplier of $\mathcal{Y}_\varepsilon^\pm|_{\Omega^{\pm}}$, respectively. Now we define the piecewise $C^1$ function $W$ with respect to the curve $\gamma \equiv \{y=0\}$ as $W|_{\Omega^{\pm}} = V^\pm$. Notice that $\gamma$ is an invariant curve of $\mathcal{Y}_\varepsilon$ in agreement with Theorem \ref{Teo-inv}. According to Remark \ref{rem}, $W$ is a weak inverse Jacobi multiplier of $\mathcal{Y}_\varepsilon$ in $\Omega = \mathbb{R}^2$ and $1/W$ is an invariant measure of $\mathcal{Y}_\varepsilon$ in $\Omega$.

\subsection{Example: general Poisson systems}

Let us now consider an example of statement (ii) of Remark \ref{rem}. As already recalled along the article, a general planar Poisson system corresponds to the $C^\infty$ class and has the form
\begin{equation}
\label{poisson-2d}
	\frac{\mbox{\rm d}x}{\mbox{\rm d}t} = \eta(x_1,x_2) {\cal S}_{(2,2)} \cdot \nabla H (x_1,x_2)
\end{equation}
where ${\cal S}_{(2,2)}$ is the $2 \times 2$ symplectic matrix. As usual, assume the system is defined in a domain $\Omega$. The structure matrix ${\cal J}(x_1,x_2) = \eta(x_1,x_2) {\cal S}_{(2,2)}$ has constant rank 2 in $\Omega$ if and only if $\eta$ does not vanish in $\Omega$. In this case, as indicated in Remark \ref{remark13} (see also references therein) it is possible to construct the Darboux canonical form globally in $\Omega$ by means of a time reparametrization and the smooth function $\eta$ is an inverse Jacobi multiplier of the Darboux canonical form (which is now a classical Hamiltonian flow). The opposite case arises (as mentioned in item(ii) of Remark \ref{rem}) provided there is a curve $\gamma \equiv \{ (x_1, x_2) \in  \Omega : \eta (x_1,x_2)=0 \}$ leading to a partition $\Omega =\Omega_1 \cup \Omega_2 \cup \gamma$, with $\gamma =\partial \Omega_1 \cap \partial \Omega_2$, and $\eta|_{\Omega_i}$ having opposite signs. Now the global Darboux reduction is not possible since ${\cal J}$ is not regular in $\Omega$. Note that $\eta$ is still an inverse Jacobi multiplier (and now $| \eta|$ is a weak inverse Jacobi multiplier) of the Darboux canonical form in $\Omega$, and such reduction can be now carried out separately on each subdomain $\Omega_i$. From the point of view of time rescalings, in each subdomain $\Omega_i$ they will have opposite signs, namely in one subdomain the direction of time will be inverted in the Darboux reduction, while in the other it will not. Additionally, in this case $\gamma$ is an invariant curve.

\mbox{}

\mbox{}

\noindent {\bf Acknowledgements}

\noindent Both authors would like to acknowledge Ministerio de Econom\'{\i}a y Competitividad for Project Ref.
MTM2014-53703-P. In addition, I.A.G. acknowledges AGAUR grant number 2014SGR 1204. B.H.-B. acknowledges
Ministerio de Econom\'{\i}a y Competitividad for Project Ref. MTM2016-80276-P as well as financial support from Universidad Rey Juan Carlos-Banco de Santander (Excellence Group QUINANOAP, grant number 30VCPIGI14). Finally, B.H.-B. is sincerely indebted to the members of Departament de Matem\`atica, Universitat de Lleida, for their kind hospitality.


\begin{thebibliography}{99}
\bibitem{agz1} {\sc A. Ay, M. Gurses and K. Zheltukhin}, {\it Hamiltonian equations in $\mathbb{R}^3$}, J. Math. Phys. {\bf 44} (2003), 5688--5705.

\bibitem{B} {\sc N.N. Bautin}, {\it On the number of limit cycles which appear with the variation of coefficients from an equilibrium position of focus or center type}, Amer. Math. Soc. Transl. {\bf 100} (1954), 1--19.

\bibitem{BerroneGiacomini2} {\sc L.R. Berrone and H. Giacomini}, {\it \ Inverse Jacobi multipliers},  Rend. Circ. Mat. Palermo (2)
{\bf 52} (2003), 77--130.

\bibitem{rus1} {\sc I.A. Bizyaev, A.V. Borisov and I.S. Mamaev,} {\it The Hojman Construction and
Hamiltonization of Nonholonomic Systems}, Symmetry, Integrability and Geometry: Methods and Applications SIGMA {\bf 12} (2016), 1--19.

\bibitem{rus2} {\sc I.A. Bizyaev, A.V. Borisov and I.S. Mamaev,} {\it Hamiltonization of Elementary
Nonholonomic Systems}, Russian J. Math. Phys. {\bf 22} (2015), 444--453.

\bibitem{rus3} {\sc A.V. Bolsinov, A.V. Borisov and I.S. Mamaev,} {\it Geometrisation of Chaplygin's reducing
multiplier theorem}, Nonlinearity {\bf 28} (2015),  2307--2318.


\bibitem{rus4} {\sc A.V. Bolsinov, A.V. Borisov and I.S. Mamaev,} {\it Hamiltonization of Non-Holonomic
Systems in the Neighborhood of Invariant Manifolds}, Regul. Chaotic Dyn. {\bf 16} (2011),
443--464.

\bibitem{BGM} {\sc A. Buic\u{a}, I.A. Garc\' \i a and S. Maza}, {\it Inverse Jacobi multipliers: recent applications in dynamical systems}, Progress and Challenges in Dynamical Systems, Springer Proc. Math. Stat. {\bf 54} (2013) 127--141.

\bibitem{BCT} {\sc  C.A. Buzzi, T.  de Carvalho and M.A. Teixeira}, {\it Birth of limit cycles bifurcating from a nonsmooth center}, J. Math. Pures Appl. {\bf 102} (2014), 36--47.

\bibitem{CF} {\sc L. Cair\'o and M.R. Feix}, {\it On the Hamiltonian structure of 2D ODE possessing an invariant}, J. Phys. A {\bf 25} (1992), L1287--L1293.

\bibitem{CS} {\sc J. Chavarriga and M. Sabatini}, {\it A survey of isochronous centers}, Qual. Theory Dyn. Syst. {\bf 1} (1999), 1--70.

\bibitem{BBCK} {\sc M. di Bernardo, C.J. Budd, A.R. Champneys and P. Kowalczyk}, {\it Piecewise-Smooth Dynamical Systems, Theory and Applications}, Springer-Verlag, London, 2008.

\bibitem{F} {\sc A.F. Filippov}, {\it Differential Equation with Discontinuous Right-Hand Sides}, Kluwer Academic, Netherlands, 1988.

\bibitem{Isaac}  {\sc I.A. Garc\' \i a}, {\it Integrable zero-Hopf singularities and 3-dimensional centers}. To appear in Proc. Roy. Soc. Edinburgh Sect. A.

\bibitem{dlrjc1} {\sc I.A. Garc\'{\i}a and B. Hern\'{a}ndez-Bermejo}, {\it Perturbed Euler top and bifurcation of limit cycles on invariant Casimir surfaces}, Physica D {\bf 239} (2010), 1665--1669.

\bibitem{dlrjc2} {\sc I.A. Garc\'{\i}a and B. Hern\'{a}ndez-Bermejo}, {\it Poisson systems as the natural framework for additional first integrals via Darboux invariant hypersurfaces}, Bull. Sci. Math. {\bf 137} (2013), 242--250.

\bibitem{dlrjc3} {\sc I.A. Garc\'{\i}a and B. Hern\'{a}ndez-Bermejo}, {\it Periodic orbits in analytically perturbed Poisson systems}, Physica D {\bf 276} (2014), 1--6.

\bibitem{GV} {\sc I.A. Garc\' \i a and C. Valls}, {\it The three-dimensional center problem for the zero-Hopf singularity}, Discrete Contin. Dyn. Syst. {\bf 36} (2016), 2027--2046.

\bibitem{bs6} {\sc B. Hern\'{a}ndez-Bermejo}, {\it Generalization of solutions of the Jacobi PDEs associated to time reparametrizations of Poisson systems}, J. Math. Anal. Appl. {\bf 344} (2008), 655–-666.

\bibitem{bs7} {\sc B. Hern\'{a}ndez-Bermejo}, {\it Generalized results on the role of new-time transformations in finite-dimensional Poisson systems}, Phys. Lett. A {\bf 374} (2010), 836–-841.

\bibitem{bs8} {\sc B. Hern\'{a}ndez-Bermejo}, {\it New global solutions of the Jacobi partial differential equations}, Physica D {\bf 241} (2012), 764--774.

\bibitem{Ka1} {\sc W. Kapteyn}, {\it On the centra of the integral curves which satisfy differential equations of the
first order and the first degree}, Proc. Kon. Akad. Wet., Amsterdam {\bf 13} (1911), 1241–-1252.

\bibitem{Ka2} {\sc W. Kapteyn}, {\it New researches upon the centra of the integrals which satisfy differential equations
of the first order and the first degree}, Proc. Kon. Acad. Wet., Amsterdam {\bf 14} (1912), 1185–-1185; {\bf 15} (1912), 46-–52.

\bibitem{rus5} {\sc V.V. Kozlov,} {\it Invariant measures of smooth dynamical systems, generalized functions and summation methods}, Izvestiya: Mathematics {\bf 80} (2016), 342--358.

\bibitem{Kunze} {\sc M. Kunze}, {\it Non-smooth dynamical systems}, Springer-Verlag, Berlin, 2000.

\bibitem{L} {\sc W.S. Loud}, {\it Behavior of the period of solutions of certain plane autonomous systems near centers}, Contributions to Differential Equations {\bf 3} (1964), 21--36.

\bibitem{Nemitskii}{\sc V.V. Nemitskij and V.V. Stepanov}, {\it \ Qualitative theory of
differential equations}, Princeton, NJ: Princeton Univ. Press, 1960.

\bibitem{Nutku} {\sc Y. Nutku}, {\it Hamiltonian structure of the Lotka-Volterra equations},
Phys. Lett. {\bf 145A} (1990), 27--28.

\bibitem{olv1} {\sc P.J. Olver}, {\it Applications of Lie Groups to Differential Equations, 2nd ed.}, New York:  Springer-Verlag, 1993.

\bibitem{RS}{\sc V.G. Romanovski and D.S. Shafer}, {\it The center and cyclicity problems: a computational algebra approach}. Birkh\"auser Boston, Inc., Boston, MA, 2009.

\bibitem{S}{\sc D. Schlomiuk}, {\it Algebraic particular integrals, integrability and the problem of the center}, Trans. Amer. Math. Soc. {\bf 338} (1993), 799--841.

\bibitem{wei1} {\sc A. Weinstein}, {\it The local structure of Poisson manifolds}, J. Diff. Geom. {\bf 18} (1983), 523--557.

\bibitem{Whittaker} {\sc E.T. Whittaker}
{\it \ A treatise on the analytical dynamics of particles \& rigid bodies,
4th ed.}, Cambridge: Cambridge University Press, 1937.

\end{thebibliography}
\end{document}